\documentclass{article}
\usepackage[utf8]{inputenc}
\usepackage[english]{babel}
\usepackage{amsmath}
\usepackage{amssymb}
\usepackage[margin=1.1in]{geometry}
\usepackage{esint}
\usepackage{graphicx}
\usepackage{caption}
\usepackage{subcaption}
\usepackage{float}
\usepackage{mathtools}
\usepackage{xcolor}
\usepackage{setspace}
\usepackage[shortlabels]{enumitem}
\usepackage{MnSymbol}
\usepackage[sorting=none]{biblatex}
\usepackage{csquotes}
\numberwithin{equation}{section}
\usepackage[pdfpagelabels]{hyperref}
\usepackage{todonotes}
\usepackage{pdfpages}
\usepackage{fancyhdr}
\usepackage{dsfont}
\usepackage{authblk}

\DeclareMathOperator{\im}{Im}

\spacing{1.25}

\title{\bf Modular conjugation for multicomponent regions}

\author[1]{Nicolás Abate}
\author[1,2]{David Blanco\footnote{Email: dblanco@df.uba.ar}}
\author[1,2]{Mateo Koifman}
\author[1,2]{Guillem Pérez-Nadal\footnote{Email: guillem@df.uba.ar}}
\affil[1]{{\small Universidad de Buenos Aires, Facultad de Ciencias Exactas y Naturales, Departamento de Física. Buenos Aires, Argentina.}}
\affil[2]{{\small CONICET - Universidad de Buenos Aires, Instituto de Física de Buenos Aires (IFIBA). Buenos Aires, Argentina.}}

\date{}
\begin{document}

\maketitle

\begin{abstract}
We consider a massless Dirac field in $1+1$ dimensions, and compute the Tomita-Takesaki modular conjugation corresponding to the vacuum state and a generic multicomponent spacetime region. We do it by analytic continuation from the modular flow, which was computed recently. We use our result to discuss the validity of Haag duality in this model. 
\end{abstract}

\tableofcontents
\clearpage

\pagenumbering{arabic}
\vspace*{1cm}

\section{Introduction}

In the last years, 
a lot has been learned about quantum field theory (QFT) by studying its entanglement properties. One famous example is the irreversibility of the renormalization group flow which was proved in various dimensions using the strong subadditity property of entanglement entropy \cite{Casini:2006es,Casini:2012ei,Casini:2015woa,Casini:2017vbe}. There have also been many applications to holography (see \cite{VanRaamsdonk:2016exw} for a review) and, more recently, to the black hole information problem \cite{Penington:2019npb,Almheiri:2019psf,Almheiri:2020cfm}.

Usually, entanglement is characterized in terms of the reduced density matrix. However, this object is not well-defined in QFT because the local algebras of operators do not admit a trace.
The usual way to circumvent this problem is to put the QFT on a lattice in order to obtain a discrete set of degrees of freedom which allows for the definition of a trace. Then, after the computations are done, one 
sends the lattice spacing to zero
to restore the relativistic symmetry, retains the information that survives
the limit (and that is independent of the details of the regularization scheme used) and interprets it as a property
of the QFT. This has proven to be a very successful approach and most of the prominent results in the area were derived with this idea behind.

It is interesting, though, that there are objects intimately related to entanglement 
which are well-defined in QFT.
Two of these 
objects
appear in the context of Tomita-Takesaki theory (see section \ref{section2}) and are known as the modular operator ($\Delta$) and the modular conjugation ($J$); they essentially play the role of the modulus and phase in the polar decomposition of an operator called the Tomita operator. The study of these 
objects
is very relevant since they 
carry information about entanglement and they are directly accessible in the QFT.

The modular operator $\Delta$ is related to the 
reduced density matrix, or equivalently to its logarithm, the modular Hamiltonian.
Among many 
applications in QFT, the knowledge of modular Hamiltonians was crucial for the formulation of a well-defined version of the Bekenstein bound \cite{Casini:2008cr} and for the proof of several energy inequalities \cite{Blanco:2013lea,Faulkner:2016mzt,Blanco:2017akw,Balakrishnan:2017bjg}. Modular Hamiltonians also found applications in the context of holography, for instance in the derivation of the linearized Einstein equations in the bulk from entanglement properties of the boundary CFT \cite{Faulkner:2013ica,Lashkari:2013koa,Blanco:2018riw,Swingle:2014uza}. These objects have been computed for several regions and states. The first results of modular Hamiltonians were local \cite{Bisognano:1976za,Unruh:1976db,Casini:2011kv,Cardy:2016fqc,Hartman:2015apr} but examples of non-local modular Hamiltonians were eventually derived \cite{Casini:2009vk,Wong:2013gua,Arias:2018tmw,Blanco:2019xwi,Blanco:2019cet,Fries:2019ozf}.

The modular conjugation $J$ can be used to understand the structure of the local algebras of the QFT. A general feature of QFT is that operators localized in spacelike separated regions commute{\footnote{Of course, this statement has to be modified if there are fermionic operators, we will come to this point below.}}. In other words, given a region ${\mathcal U}$, any operator localized in a spacelike separated region will commute with everything in ${\mathcal U}$. But there may be other operators, not localized in a spacelike separated region, which commute with everything in ${\mathcal U}$. If there are not, 
one says that Haag duality holds. Haag duality is known to hold in some cases \cite{araki1964neumann,Bisognano:1976za} and to fail in others \cite{Haag:1992hx}; the general conditions under which it holds or fails are an open question. The modular conjugation is useful in this context because it can be used to determine the commutant of a local algebra, i.e., the set of all operators commuting with everything in a given region.

In contrast with the modular operator, the modular conjugation has not been studied so extensively and it is only known in very few cases. When the global state is the vacuum and the region is the Rindler wedge, the modular conjugation is essentially the CPT operator for any QFT \cite{Bisognano:1976za}. Using this result and conformal transformations one can obtain the modular conjugation for any CFT when the region is a causal diamond and the global state is the vacuum (this was done for massless scalar fields in \cite{Hislop:1981uh}). All the mentioned results up to this point involve regions with only one component. In this paper we obtain a new modular conjugation, namely that corresponding to the free massless Dirac field in $1+1$ dimensions in the vacuum state for generic multicomponent regions.

The paper is organized as follows. In section \ref{section2} we 
introduce the modular operator $\Delta$ and the modular conjugation $J$ and discuss their properties. 
In section \ref{section3} we describe the relevant aspects of the model we will consider, the massless Dirac field in $1+1$ dimensions.
In section \ref{s4} we review the computation of the modular conjugation of the Rindler wedge and the causal diamond; we particularize this computation to the massless fermion in $1+1$,
with emphasis on the subtleties
that arise when one considers fermions instead of bosons. In section \ref{section6}, we compute $J$ for a generic multicomponent region, and use our result to discuss the validity of Haag duality in this model.
Finally, we conclude 
with a discussion in section \ref{section7}.

\section{Modular operator and modular conjugation}\label{section2}

Let ${\mathcal H}$ be a Hilbert space, and let ${\mathcal B}({\mathcal H})$ be the algebra of all bounded operators on ${\mathcal H}$. Given any set ${\mathcal S}\subseteq{\mathcal B}({\mathcal H})$, the commutant ${\mathcal S}'$, namely the set of all operators in ${\mathcal B}({\mathcal H})$ which commute with all operators in ${\mathcal S}$, is clearly an algebra containing the identity. Moreover, if ${\mathcal S}$ is self-adjoint (i.e., closed under the operation of taking adjoints) then so is ${\mathcal S}'$. The commutants of self-adjoint sets are called {\emph{von Neumann algebras}}. Note that, if ${\mathcal S}\subseteq{\mathcal T}$, then ${\mathcal T}'\subseteq{\mathcal S}'$, and that ${\mathcal S}\subseteq{\mathcal S}''$. From these two properties it follows that, if ${\mathcal S}\subseteq{\mathcal T}'$, then ${\mathcal S}''\subseteq{\mathcal T}'$, so ${\mathcal S}''$ is the smallest commutant containing ${\mathcal S}$; in particular, ${\mathcal S}''={\mathcal S}$ if ${\mathcal S}$ is itself a commutant. Therefore, the bicommutant of a self-adjoint set is the smallest von Neumann algebra containing that set, and, in particular, every von Neumann algebra is equal to its bicommutant.

Let ${\mathcal A}\subseteq{\mathcal B}({\mathcal H})$ be a von Neumann algebra. A vector $|\Omega\rangle\in{\mathcal H}$ is said to be {\emph{cyclic}} for ${\mathcal A}$ if the subspace ${\mathcal A}|\Omega\rangle$ is dense in ${\mathcal H}$, meaning that any vector in the Hilbert space can be approximated arbitrarily well by an element of this subspace; $|\Omega\rangle$ is said to be {\emph{separating}} for ${\mathcal A}$ if the condition $a\in{\mathcal A}$, $a|\Omega\rangle=0$ implies $a=0$, or, in other words, 
if the map $a\mapsto a|\Omega\rangle$ from the algebra to the Hilbert space is one to one.
Note that, if $|\Omega\rangle$ is cyclic for ${\mathcal A}$, then it is separating for ${\mathcal A}'$: indeed, the condition $a'\in{\mathcal A}'$, $a'|\Omega\rangle=0$ implies $a'{\mathcal A}|\Omega\rangle=0$ and hence, by continuity, $a'=0$ if ${\mathcal A}|\Omega\rangle$ is dense. In fact, one can show (see e.g.~\cite{Witten:2018zxz}) that the converse is also true: if $|\Omega\rangle$ is separating for ${\mathcal A}'$, then it is cyclic for ${\mathcal A}$. Therefore, cyclic for an algebra is the same as separating for its commutant.

We are interested in von Neumann algebras admitting a cyclic and separating vector. The study of such algebras is called Tomita-Takesaki theory (see \cite{Witten:2018zxz,Summers:2003tf,Haag:1992hx} for physics-oriented reviews and \cite{takesaki,kadison1986fundamentals} for more detailed treatments).
Suppose that $|\Omega\rangle$ is cyclic and separating for ${\mathcal A}$, and consider the operator $S$ defined on ${\mathcal A}|\Omega\rangle$ by
\begin{equation}\label{tomita}
    Sa|\Omega\rangle=a^\dagger|\Omega\rangle.
\end{equation}
This is called the Tomita operator associated with ${\mathcal A}$ and $|\Omega\rangle$. Note that the separating property ensures that this definition makes sense; the cyclic property implies that $S$ is densely defined. Clearly, $S$ is an antilinear operator satisfying $S^2={\mathds 1}$ (hence invertible) and $S|\Omega\rangle=|\Omega\rangle$; it also satisfies $S^\dagger|\Omega\rangle=|\Omega\rangle$ (recall that the adjoint of an antilinear operator is defined by $\langle\psi|O^\dagger|\phi\rangle=\langle\phi|O|\psi\rangle$), because $\langle a\Omega|S^\dagger\Omega\rangle=\langle\Omega|Sa\Omega\rangle=\langle\Omega|a^\dagger\Omega\rangle=\langle a\Omega|\Omega\rangle$.
As it turns out, $S$ is unbounded but closable{\footnote{An operator $O$ on ${\mathcal H}$ is said to be closed if its graph is a closed subspace of ${\mathcal H}\oplus{\mathcal H}$; it is said to be closable if the closure of its graph is the graph of some operator, which is then called the closure of $O$.}}; its domain is slightly extended beyond ${\mathcal A}|\Omega\rangle$ by taking the closure. This makes $S$ an invertible, densely defined closed operator, which in turn guarantees that it admits a unique polar decomposition,
\begin{equation}
    S=J\Delta^{1/2}
\end{equation}
with $\Delta$ positive and $J$ antiunitary (note that $\Delta=S^\dagger S)$. These operators are called respectively the {\emph{modular operator}} and the {\emph{modular conjugation}} associated with ${\mathcal A}$ and $|\Omega\rangle$.

Let us discuss the properties of the modular objects. Since $S^2={\mathds 1}$, we have $J\Delta^{1/2}J=\Delta^{-1/2}$, which after some algebra implies $J^\dagger=J$ and hence $J^2={\mathds 1}$. On the other hand, since $S|\Omega\rangle=S^\dagger|\Omega\rangle=|\Omega\rangle$ we have $\Delta|\Omega\rangle=J|\Omega\rangle=|\Omega\rangle$. Another property, which is highly non-trivial, is
\begin{equation}
    \Delta^{is}{\mathcal A}\Delta^{-is}={\mathcal A}\quad(s\in{\mathbb R})\qquad J{\mathcal A}J={\mathcal A}'.
\end{equation}
Thus, the modular operator defines a one-parameter group of automorphisms of ${\mathcal A}$, which is called the {\emph{modular flow}}, and the modular conjugation defines an antilinear isomorphism between ${\mathcal A}$ and its commutant. This is regarded as the main result of Tomita-Takesaki theory.

The modular flow can also be characterized by the so-called modular condition. Let $\alpha:{\mathbb R}\times{\mathcal A}\to{\mathcal A}$ be a one-parameter group of automorphisms of ${\mathcal A}$, that to each $s\in{\mathbb R}$ assigns the automorphism $\alpha_s$. One says that $\alpha$ satisfies the {\emph{modular condition}} with respect to $|\Omega\rangle$ if, for every $a,b\in{\mathcal A}$, there is a complex function $G(z)$ on the strip $\im z\in[-1,0]$, analytic on the interior of the strip and continuous on its boundary, such that, for $s\in{\mathbb R}$,
\begin{equation}\label{kms}
    G(s)=\langle\Omega|a\alpha_s(b)|\Omega\rangle\qquad G(s-i)=\langle\Omega|b\alpha_{-s}(a)|\Omega\rangle.
\end{equation}
As it turns out, there is a unique one-parameter group of automorphisms satisfying the modular condition, and this is the modular flow $\alpha_s(a)=\Delta^{is}a\Delta^{-is}$. An intermediate result which is used to show this, and which we will use extensively in the following, is that, for $a\in{\mathcal A}$, the map $z\mapsto\Delta^{iz}a|\Omega\rangle$ is analytic on the interior of the strip $\im z\in[-1/2,0]$ and continuous on its boundary.

In order to gain familiarity with the modular objects, let us see what they look like in a simple finite-dimensional example. Suppose that ${\mathcal H}={\mathcal H}_1\otimes{\mathcal H}_2$, with ${\mathcal H}_1$ and ${\mathcal H}_2$ of the same dimension $n$, and consider the algebra ${\mathcal A}={\mathcal B}({\mathcal H}_1)\otimes{\mathds 1}_2$, which is a von Neumann algebra. For any state vector $|\Omega\rangle\in{\mathcal H}$, there are orthonormal bases $\{|i\rangle_1\}\subset{\mathcal H}_1$ and $\{|i\rangle_2\}\subset{\mathcal H}_2$ and a probability distribution $p_1,\dots,p_n$ such that
\begin{equation}
    |\Omega\rangle=\sum_{i=1}^n\sqrt{p_i}|i\,i\rangle.
\end{equation}
This is the Schmidt decomposition of $|\Omega\rangle$. Note that $|i\rangle_1$ and $|i\rangle_2$ are eigenvectors of the reduced density matrices $\rho_1$ and $\rho_2$ respectively, both with the same eigenvalue $p_i$. A necessary and sufficient condition for $|\Omega\rangle$ to be cyclic and separating for ${\mathcal A}$ is that all these probabilities be non-vanishing or, in other words, that both reduced density matrices be invertible. Suppose that this is the case. Setting $a=(|i\rangle\langle j|)_1\otimes{\mathds 1}_2$ in the definition of the Tomita operator, Eq.~(\ref{tomita}), one finds that $S|i\,j\rangle=\sqrt{p_i/p_j}\,|j\,i\rangle$ and therefore
\begin{equation}\label{DeltaRho}
    \Delta=\rho_1\otimes\rho_{2}^{-1}\qquad J|i\,j\rangle=|j\,i\rangle.
\end{equation}
It is a simple matter to check that $\Delta$ and $J$ satisfy all the properties discussed above. As we see, knowing $\Delta$ is the same as knowing the reduced density matrices. Of course, these do not contain all the information about the global state $|\Omega\rangle$; the remaining information is contained in $J$.

In QFT, every open spacetime region ${\mathcal U}$ has naturally associated a von Neumann algebra, namely the algebra ${\mathcal A}_{\mathcal U}$ generated by the bounded operators localized in ${\mathcal U}$ (i.e., the bicommutant of that set of operators). In the algebraic approach to QFT, this assignment of algebras to regions is viewed as the essential feature of a QFT. Note that it satisfies the following properties: (i) if ${\mathcal U}\subseteq{\mathcal V}$, then ${\mathcal A}_{\mathcal U}\subseteq{\mathcal A}_{\mathcal V}$, and (ii) if ${\mathcal U}$ and ${\mathcal V}$ are spacelike separated, then ${\mathcal A}_{\mathcal U}\subseteq{\mathcal A}_{\mathcal V}'$ (assuming that the algebras contain no fermionic operators; otherwise this property has to be suitably modified, as we will discuss below). A famous result in QFT known as the Reeh-Schlieder theorem establishes that the vacuum $|\Omega\rangle$ is cyclic for ${\mathcal A}_{\mathcal U}$ provided only that ${\mathcal U}$ is non-empty (it can otherwise be arbitrarily small). If the causal complement ${\mathcal U}'$, i.e., the largest open region spacelike separated from ${\mathcal U}$, is also non-empty, then $|\Omega\rangle$ is also cyclic for ${\mathcal A}_{{\mathcal U}'}$, hence separating for ${\mathcal A}_{{\mathcal U}'}'$ and in particular for ${\mathcal A}_{\mathcal U}\subseteq{\mathcal A}_{{\mathcal U}'}'$. Thus, the Reeh-Schlieder theorem implies that the vacuum is cyclic and separating for ${\mathcal A}_{\mathcal U}$ provided that both ${\mathcal U}$ and ${\mathcal U}'$ are non-empty. In the case where ${\mathcal U}$ is the Rindler wedge, the modular objects are known for any QFT. This is another famous result, called the Bisognano-Wichmann theorem. The modular operator in this case is related to a boost generator, and the modular conjugation is essentially the CPT operator. From the modular operator one learns that uniformly accelerated observers see the vacuum as a thermal state (the Unruh effect); from the modular conjugation one learns that, if ${\mathcal U}$ is the Rindler wedge, then ${\mathcal A}_{\mathcal U}'={\mathcal A}_{{\mathcal U}'}$. This property (that there is nothing else in the commutant than what can be found in the causal complement) 
is known as {\emph{Haag duality}}, and the precise conditions under which it holds are an open question. Given the amount of information one extracts from the modular objects in the case of the Rindler wedge, it is natural to ask what these objects look like for more general regions. The modular operator is known in a few cases, the modular conjugation has been less explored. Contributing to fill this gap is the purpose of this paper.

\section{Massless fermion in 1+1}\label{section3}

\begin{figure}
    \centering
    \captionsetup{justification=centering}
    \includegraphics[scale=0.6]{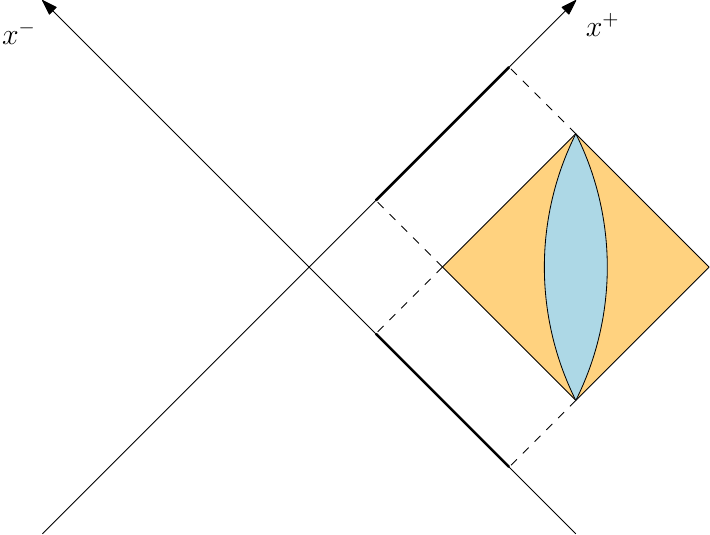}
    \caption{Two spacetime regions with the same null projections and hence the same algebra.}
    \label{figprojections}
\end{figure}

We focus on a very simple QFT, namely that of a massless Dirac field in $(1+1)$-dimensional Minkowski spacetime. In $1+1$ dimensions, the Dirac field is a two-component spinor, $\Psi=(\Psi_+,\Psi_-)$. The massless equation of motion then implies that each component (chirality) is a function of a single null coordinate, $\Psi_\pm(x^+,x^-)=\psi_\pm(x^\pm)$. These functions are subject to the canonical anticommutation relations,
\begin{equation}\label{car}
    \{\psi_\pm(x),\psi_{\pm}^\dagger(y)\}=\delta(x-y),
\end{equation}
the remaining anticommutators being zero. Due to the delta function above, $\psi_\pm$ is not really a function but a distribution, which has to be smeared with a test function $f$ to give a well-defined operator, $\psi_\pm(f)=\int_{-\infty}^\infty dx\,\psi_\pm(x)f(x)$. Note that this operator is bounded because, by (\ref{car}),
\begin{equation}
    [\psi_{\pm}(f)]^\dagger\psi_\pm(f)=\|f\|^2-\psi_\pm(f)[\psi_{\pm}(f)]^\dagger\le \|f\|^2,
\end{equation}
where $\|f\|^2=\int_{-\infty}^\infty dx\,|f(x)|^2$, which is finite for any test function. The local algebra associated with an open spacetime region ${\mathcal U}$ is generated by the smearings of $\psi_+$, $\psi_-$ and their adjoints with all test functions supported in the corresponding null projections of ${\mathcal U}$,
\begin{equation}
    {\mathcal A}_{\mathcal U}=\{\psi_+(f_+),\psi_{+}^\dagger(f_+),\psi_-(f_-),\psi_{-}^\dagger(f_-),{\text{ supp}}(f_\pm)\subseteq\pi^\pm({\mathcal U})\}'',
\end{equation}
where $\pi^\pm(x^+,x^-)=x^\pm$ is the projection onto the $x^\pm$ axis. Note that two regions ${\mathcal U}$ and ${\mathcal V}$ with the same projections onto the null axes have the same algebra: if $\pi^\pm({\mathcal U})=\pi^\pm({\mathcal V})$, then ${\mathcal A}_{\mathcal U}={\mathcal A}_{\mathcal V}$. An example of two such regions is given in figure \ref{figprojections}. This will be important when we discuss Haag duality below. 
The vacuum state $|\Omega\rangle$ is Gaussian with
\begin{equation}\label{2pt}
    \langle\Omega|\psi_\pm(x)\psi_{\pm}^\dagger(y)|\Omega\rangle=\langle\Omega|\psi_{\pm}^\dagger(x)\psi_\pm(y)|\Omega\rangle=\frac{1}{2\pi i}\,\frac{1}{x-y-i\epsilon},
\end{equation}
where $\epsilon>0$ is to be sent to zero after smearing; the remaining two-point functions all vanish. Note that the right-hand side above can be extended continuously in $y$ to an analytic function in the upper half-plane (positive imaginary part),
so the same is true for $\psi_\pm(y)|\Omega\rangle$ and $\psi_{\pm}^\dagger(y)|\Omega\rangle${\footnote{More precisely, for any test function $f$ the vector-valued functions $\psi_\pm(f_y)|\Omega\rangle$ and $\psi^{\dagger}_\pm(f_y)|\Omega\rangle$, where $f_y(x)=f(x-y)$, can be extended continuously in $y$ to analytic functions in the upper half-plane.
}}. This property will be important in what follows.

Let us discuss the symmetries of this model, i.e., the transformations $\psi_\pm\to\psi_{\pm}'$ which preserve the canonical anticommutation relations (\ref{car}). Any such transformation defines an automorphism $\alpha$ of the algebra ${\mathcal A}$ spanned by the identity and all products of smeared fields by the equation $\alpha(\psi_\pm(f))=\psi_\pm'(f)$.
We are interested in transformations which also preserve the two-point functions (\ref{2pt}), whose corresponding automorphisms satisfy $\langle\Omega|\alpha(a)|\Omega\rangle=\langle\Omega|a|\Omega\rangle$ for all $a\in{\mathcal A}$.
These transformations are implemented by unitary operators which leave the vacuum invariant. Indeed, the above property of $\alpha$ enables us to define an operator $U$ by the equation $U a|\Omega\rangle=\alpha(a)|\Omega\rangle$; one can easily check that this operator is unitary, leaves $|\Omega\rangle$ invariant and satisfies $U aU^\dagger=\alpha(a)$ for all $a\in{\mathcal A}$. A transformation preserving (\ref{car}) and (\ref{2pt}) (and thus implemented by a unitary operator which leaves the vacuum invariant) is
\begin{equation}\label{conf}
    \psi_\pm'(x)=\frac{1}{c_\pm x+d_\pm}\psi_\pm((a_\pm x+b_\pm)/(c_\pm x+d_\pm))
\end{equation}
with $a_\pm d_\pm-b_\pm c_\pm=1$. Note that the associated spacetime transformation is a conformal transformation. The particular case $b_\pm=c_\pm=0$, $a_\pm=1/d_\pm=e^{\pm\eta/2}$,
\begin{equation}\label{boost}
    \psi_\pm'(x)=e^{\pm\eta/2}\psi_\pm(e^{\pm\eta}x),
\end{equation}
corresponds to a boost of parameter $\eta$. Other transformations preserving (\ref{car}) and (\ref{2pt}) are the $U(1)$ and charge conjugation transformations,
\begin{equation}\label{u1c}
    \psi_\pm'=e^{i\theta_\pm}\psi_\pm\qquad\psi_\pm'=\psi_\pm^\dagger
\end{equation}
with $\theta_\pm\in[0,2\pi)$. On the other hand, the PT transformation
\begin{equation}\label{pt}
    \psi_{\pm}'(x)=\psi_\pm(-x)
\end{equation}
preserves the anticommutation relations (\ref{car}) but not the two-point functions (\ref{2pt}), which are mapped to their complex conjugates. Transformations with this property are implemented by antiunitary (rather than unitary) operators which leave the vacuum invariant, as can be shown by exactly the same construction as above.

In this QFT, as is always the case with fermions, field operators localized in two spacelike separated regions ${\mathcal U}$ and ${\mathcal V}$ anticommute instead of commuting. Of course, this does not mean that the corresponding algebras anticommute: for example, the product of two field operators in ${\mathcal U}$ commutes with everything in ${\mathcal V}$. What is, then, the relation between the algebras ${\mathcal A}_{\mathcal U}$ and ${\mathcal A}_{\mathcal V}$? To answer this question, consider the unitary operator $U_\pi$ associated with the $U(1)$ transformation (first equation in (\ref{u1c})) with $\theta_+=\theta_-=\pi$, 
\begin{equation}\label{upi}
    U_\pi\psi_\pm U_{\pi}^\dagger=-\psi_\pm.
\end{equation}
Clearly, this operator anticommutes with $\psi_\pm$ and $\psi_{\pm}^\dagger$, which implies that the product $X U_\pi$, with $X$ a field operator localized in ${\mathcal U}$, commutes with everything in ${\mathcal V}$. Note also that $U_{\pi}^2={\mathds 1}$ and hence $U_{\pi}^\dagger=U_\pi$. Taking this into account one finds that the operator
\begin{equation}\label{z}
    Z=\frac{{\mathds 1}+iU_\pi}{1+i},
\end{equation}
which is called the twist operator,
is unitary and satisfies
\begin{equation}\label{zpsi}
    Z\psi_\pm Z^\dagger=-i\psi_\pm U_\pi.
\end{equation}
Therefore, $ZXZ^\dagger$ commutes with everything in ${\mathcal V}$. In other words, if ${\mathcal S}_{\mathcal U}$ denotes the set of field operators localized in ${\mathcal U}$ (so that ${\mathcal A}_{\mathcal U}={\mathcal S}_{\mathcal U}''$), we have $Z{\mathcal S}_{\mathcal U}Z^\dagger\subseteq{\mathcal S}_{\mathcal V}'={\mathcal A}_{\mathcal V}'$. For any set ${\mathcal S}$ and any unitary operator $U$ it is easy to see that $(U{\mathcal S}U^\dagger)'=U{\mathcal S}'U^\dagger$, so this inclusion can be promoted to a relation between von Neumann algebras,
\begin{equation}\label{asspacelike}
    Z{\mathcal A}_{\mathcal U}Z^\dagger\subseteq{\mathcal A}_{\mathcal V}'.
\end{equation}
This is the relation we were looking for. Note that $Z^\dagger\psi_\pm Z=-Z\psi_\pm Z^\dagger$ and hence $Z^\dagger{\mathcal A}_{\mathcal U}Z=Z{\mathcal A}_{\mathcal U}Z^\dagger$, so the above equation is equivalent to $Z{\mathcal A}_{\mathcal V}Z^\dagger\subseteq{\mathcal A}_{\mathcal U}'$, as it should be because the regions ${\mathcal U}$ and ${\mathcal V}$ are arbitrary. 
We will refer to $Z{\mathcal A}_{\mathcal U}Z^\dagger$ as the twisted algebra of ${\mathcal U}$. 
Note from (\ref{z}) that $Z$ leaves the vacuum invariant, because so does $U_\pi$. This guarantees that the corollary we discussed above of the Reeh-Schlieder theorem (that the vacuum is cyclic and separating for non-empty regions with a non-empty causal complement) remains true in the presence of fermions.
In the context of this or any other fermionic model, we will say that Haag duality holds for a region ${\mathcal U}$ if ${\mathcal A}_{\mathcal U}'=Z{\mathcal A}_{{\mathcal U}'}Z^\dagger$, i.e., if there is nothing else in the commutant than what can be found in the twisted algebra of the causal complement. This is sometimes called a twisted version of Haag duality.

\begin{figure}
    \centering
    \captionsetup{justification=centering}
    \includegraphics[scale=0.6]{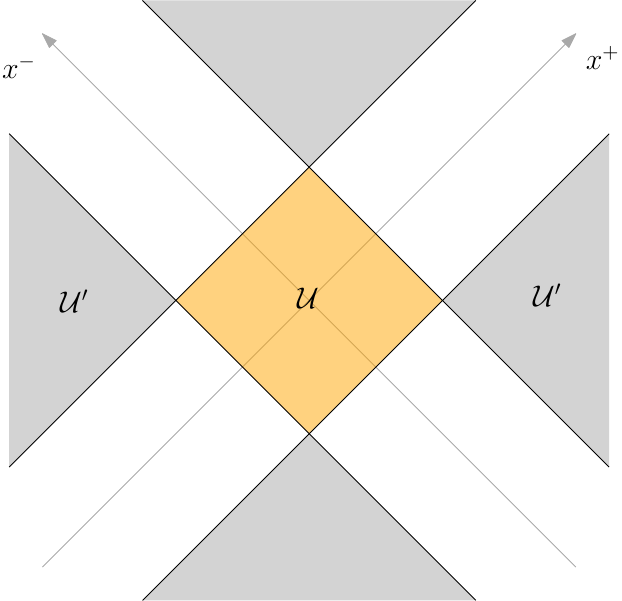}
    \caption{A causal diamond ${\mathcal U}$ and its null complement ${\mathcal U}^*$, i.e., the largest open set of points which cannot be joined with ${\mathcal U}$ by a light ray (gray region). Two of the four components of ${\mathcal U}^*$ form the causal complement ${\mathcal U}'$, which has the same null projections as ${\mathcal U}^*$.}
    \label{figumonio}
\end{figure}

In fact, Eq.~(\ref{car}) tells us that field operators localized in ${\mathcal U}$ and ${\mathcal V}$ anticommute (and hence the algebra of one region commutes with the twisted algebra of the other)
whenever there is no light ray joining these regions, which is a weaker condition than the condition of being spacelike separated. The largest open set of points that cannot be joined with a region ${\mathcal U}$ by a light ray will be referred to as the {\emph{null complement}} of ${\mathcal U}$, and will be denoted as ${\mathcal U}^*$. We thus have $Z{\mathcal A}_{{\mathcal U}^*}Z^\dagger\subseteq{\mathcal A}_{\mathcal U}'$ in this model. Since ${\mathcal U}^*$ is larger than ${\mathcal U}'$, at first sight this might seem to imply a violation of Haag duality, but this is not necessarily the case: for example, if ${\mathcal U}$ is a causal diamond as in figure \ref{figumonio}, the projections of ${\mathcal U}^*$ onto the null axes coincide with those of ${\mathcal U}'$ and hence, as discussed above, ${\mathcal A}_{{\mathcal U}^*}={\mathcal A}_{{\mathcal U}'}$. 

\section{Modular conjugation for single-component regions}\label{s4}

\subsection{Rindler wedge}\label{section4}

We now show the explicit calculation of the modular conjugation for the vacuum state and the algebra of the Rindler wedge, first derived by Bisognano and Wichmann \cite{Bisognano:1976za}. This is an extremely important result since it is an explicit example which holds for any QFT and shows the validity of Haag duality for this region. 

We restrict to the massless fermion in $1+1$,
although the same ideas are readily generalized to arbitrary QFTs in any number of dimensions. The Rindler wedge ${\mathcal R}$ is the set of all points $(x^+,x^-)$ with $x^+>0$ and $x^-<0$. The corresponding modular flow is given by
\begin{equation}\label{modflow}
    \Delta^{is}\psi_\pm(x^\pm)\Delta^{-is}=e^{\mp\pi s}\psi_\pm\left(e^{\mp 2\pi s}x^\pm\right)\equiv\alpha_s\left(\psi_\pm(x^\pm)\right),
\end{equation}
which, by comparison with (\ref{boost}), tells us that $\Delta^{is}$ is the unitary operator associated with a boost of parameter $-2\pi s$. One way to convince oneself that the above equation is correct is to verify that $\alpha$ is a one-parameter group of automorphisms of ${\mathcal A}_{\mathcal R}$ (which it is, because the boosts map the Rindler wedge to itself) and that it satisfies the modular condition. It does:
for $(x^+,x^-),(y^+,y^-)\in{\mathcal R}$, the function
\begin{equation}
    G(z)=\frac{1}{2\pi i}\frac{1}{e^{\pm\pi z}x^\pm-e^{\mp\pi z}y^\pm-i\epsilon}
\end{equation}
is analytic on the interior of the strip $\im z\in[-1,0]$ and continuous on its boundary, and satisfies the boundary conditions (\ref{kms}) with $a=\psi_\pm(x^\pm)$ and $b=\psi_{\pm}^\dagger(y^\pm)$, and also with $a=\psi_{\pm}^\dagger(x^\pm)$ and $b=\psi_\pm(y^\pm)$. That the modular condition is satisfied for arbitrary choices of $a$ and $b$ just follows by Gaussianity.

The modular flow can be used to determine the modular conjugation. Indeed, the modular objects associated with an algebra ${\mathcal A}$ and a cyclic and separating vector $|\Omega\rangle$ are related by
\begin{equation}\label{JDelta}
    Ja|\Omega\rangle=JSa^\dagger|\Omega\rangle=\Delta^{1/2}a^\dagger|\Omega\rangle
\end{equation}
for all $a\in{\mathcal A}$, where in the first step we have used the definition of the Tomita operator and in the second step we have used the property $J^2={\mathds 1}$. Moreover, the right-hand side can be obtained by analytic continuation from the modular flow. In our case we have
\begin{equation}
    \Delta^{is}\psi_{\pm}^\dagger(x^\pm)|\Omega\rangle=e^{\mp\pi s}\psi_{\pm}^\dagger\left(e^{\mp 2\pi s}x^\pm\right)|\Omega\rangle.
\end{equation}
We know from section \ref{section2} that the left-hand side is analytic in $s$ on the interior of the strip $\im s\in[-1/2,0]$ and continuous on its boundary. The same is true for the right-hand side, because for $s$ in that strip the argument of $\psi_{\pm}^\dagger$ has a non-negative imaginary part (recall the discussion under Eq.~(\ref{2pt})). Therefore, by the uniqueness of the analytic continuation, the above equation remains valid everywhere in the strip, and in particular for $s=-i/2$,
\begin{equation}\label{delta12rind}
    \Delta^{1/2}\psi_{\pm}^\dagger(x^\pm)|\Omega\rangle=\pm i\psi_{\pm}^\dagger(-x^\pm)|\Omega\rangle.
\end{equation}
Substituting into (\ref{JDelta}) with $a=\psi_\pm(x^\pm)$ we obtain
\begin{equation}
    J\psi_\pm(x^\pm)|\Omega\rangle=\pm i\psi_{\pm}^\dagger(-x^\pm)|\Omega\rangle.
\end{equation}
Since both $J$ and $U_\pi$ (the unitary operator associated with the symmetry $\psi_\pm\to-\psi_\pm$ that appeared at the end of last section) leave the vacuum invariant, this equation can be rewritten using (\ref{zpsi}) in the form
\begin{equation}
    \left[J\psi_\pm(x^\pm)J\pm Z\psi_{\pm}^\dagger(-x^\pm)Z^\dagger\right]|\Omega\rangle=0.
\end{equation}
Now, we know from Tomita-Takesaki theory that the first term between square brackets lies in ${\mathcal A}_{\mathcal R}'$. The same is true for the second, because $-(x^+,x^-)\in{\mathcal R}'$. Hence, the entire operator between square brackets is in ${\mathcal A}_{\mathcal R}'$. But the vacuum is separating for this algebra, so the operator must vanish,
\begin{equation}\label{Jpsi}
    J\psi_{\pm}(x^\pm)J=\mp Z\psi^\dagger(-x^\pm)Z^\dagger.
\end{equation}
In other words,
\begin{equation}\label{Jrindler}
    J=Z\Theta,
\end{equation}
where $\Theta$ is the antiunitary operator associated with the CPT symmetry $\psi_\pm(x)\to\mp\psi_{\pm}^\dagger(-x)$.
As a consistency check, note that $\Theta^2={\mathds 1}$ and that $\Theta$ commutes with
%$\Theta^2={\mathds 1}$ (or, in other words, $\Theta$ is self-adjoint) and that $\Theta$ commutes with 
$U_\pi$, which, by antilinearity, implies $\Theta Z=Z^\dagger\Theta$. We thus have $J^2=Z\Theta Z\Theta=ZZ^\dagger\Theta^2={\mathds 1}$, as it should be. On the other hand, note from (\ref{Jpsi}) that
\begin{equation}
    {\mathcal A}_{\mathcal R}'=J{\mathcal A}_{\mathcal R}J=Z{\mathcal A}_{{\mathcal R}'}Z^\dagger,
\end{equation}
i.e., Haag duality holds for the Rindler wedge. 

\begin{figure}
    \centering
    \captionsetup{justification=centering}
    \includegraphics[scale=0.6]{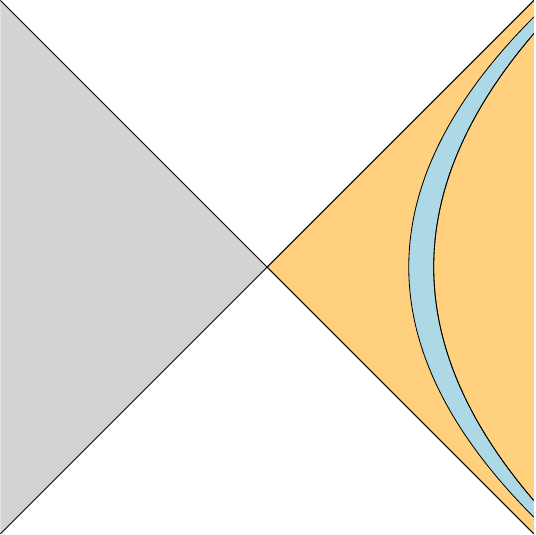}
    \caption{The Rindler wedge (orange) and a region with the same null projections (blue). Both have the same causal complement (gray region).}
    \label{figrindler}
\end{figure}

As mentioned above, these results hold in fact for any QFT. What is specific to this model is that all regions with the same null projections have the same algebra, so Eq.~(\ref{Jrindler}) gives the modular conjugation of any region with the same null projections as the Rindler wedge. An example is given in figure \ref{figrindler}. Any such region has ${\mathcal R}'$ as its causal complement, so Haag duality is satisfied for all these regions as well.

\subsection{Causal diamond}\label{section5}

In conformal field theories (CFTs), the modular conjugation associated with a causal diamond can be obtained by conformal mapping from the Rindler wedge. This was done explicitly in \cite{Hislop:1981uh} for a massless scalar field theory. Let us see how this works in the case of the massless fermion in $1+1$.

\begin{figure}
    \centering
    \captionsetup{justification=centering}
    \includegraphics[scale=0.45]{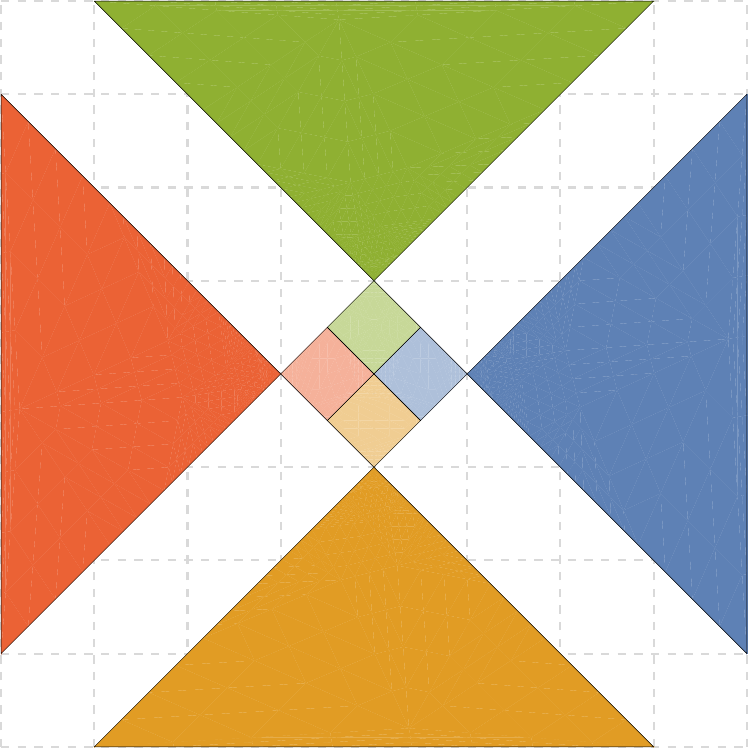}
    \caption{Regions in the casual diamond $\mathcal{D}$ (lighter colored) are mapped to regions of the corresponding darker colors outside $\mathcal{D}$ by the map $x^\pm\mapsto R^2/x^\pm$.}
    \label{Fig:J1interval}
\end{figure}

The causal diamond ${\mathcal D}$ is the set of all points $(x^+,x^-)$ with $|x^\pm|<R$ for some $R>0$. The conformal transformation
\begin{equation}\label{map}
    \sigma^\pm(x^\pm)=\pm R\frac{x^{\pm}\mp 2R}{x^{\pm} \pm 2R}
\end{equation}
maps the Rindler wedge ${\mathcal R}$ to ${\mathcal D}$: for $x^+>0$ and $x^-<0$ we have $|\sigma^\pm(x^\pm)|<R$. Hence, if $U$ is the unitary operator associated with this transformation{\footnote{More precisely, one of the unitary operators: there are four of them, because for each chirality there are two choices of the parameters $a$, $b$, $c$ and $d$ in (\ref{conf}) corresponding to the conformal transformation (\ref{map}). The unitaries do not form a representation of the conformal group but of $SL(2,{\mathbb R})\times SL(2,{\mathbb R})$, which is a covering of the conformal group. Being a fermionic theory, this should not be surprising: for example, fermions in $3$ spatial dimensions do not carry a representation of the rotation group $SO(3)$ but of its covering $SU(2)$.}},
we have ${\mathcal A}_{\mathcal D}=U{\mathcal A}_{\mathcal R}U^\dagger$. This implies a relation between the Tomita operators: for $a_{\mathcal D}\in{\mathcal A}_{\mathcal D}$ we have $a_{\mathcal R}\equiv U^\dagger a_{\mathcal D}U\in{\mathcal A}_{\mathcal R}$ and therefore
\begin{alignat}{2}
    S_{\mathcal D}a_{\mathcal D}|\Omega\rangle&=a_{\mathcal D}^\dagger|\Omega\rangle=U a_{\mathcal R}^\dagger U^\dagger|\Omega\rangle=U a_{\mathcal R}^\dagger|\Omega\rangle\nonumber\\
    &=U S_{\mathcal R}a_{\mathcal R}|\Omega\rangle=U S_{\mathcal R}U^\dagger a_{\mathcal D}U|\Omega\rangle=U S_{\mathcal R}U^\dagger a_{\mathcal D}|\Omega\rangle,
\end{alignat}
which implies
\begin{equation}
    S_{\mathcal D}=US_{\mathcal R}U^\dagger.
\end{equation}
It follows that the modular objects are also related,
\begin{equation}\label{modU}
    \Delta_{\mathcal D}=U\Delta_{\mathcal R}U^\dagger\qquad J_{\mathcal D}=UJ_{\mathcal R}U^\dagger.
\end{equation}
Together with the formula (\ref{modflow}) for the modular flow of the Rindler wedge, the first equation above gives the modular flow of the causal diamond,
\begin{eqnarray}
    &&\Delta^{is}\psi_\pm(x)\Delta^{-is}=\frac{R}{R\cosh(\pi s)-x\sinh(\pi s)}\psi_\pm(\sigma_s(x))\\
    &&\sigma_s(x)=R\frac{x\cosh(\pi s)-R\sinh(\pi s)}{R\cosh(\pi s)-x\sinh(\pi s)}.
\end{eqnarray}
Similarly, the second equation in (\ref{modU}), together with the formula (\ref{Jpsi}) for the modular conjugation of the Rindler wedge, gives the modular conjugation of the causal diamond,
\begin{equation}\label{Jdiamond}
    J\psi_\pm(x)J=\frac{R}{x}Z\psi_{\pm}^\dagger(R^2/x)Z^\dagger,
\end{equation}
where we have used that $Z$ commutes with $U$ (because so does $U_\pi$). The map $x^\pm\mapsto R^2/x^\pm$ is represented in figure \ref{Fig:J1interval}. As we see, $J$ maps the algebra of ${\mathcal D}$ to the twisted algebra of ${\mathcal D}^*$, the null complement of ${\mathcal D}$. As explained above, this algebra coincides with that of the causal complement ${\mathcal D}'$, so Haag duality is satisfied also for the causal diamond.

The remarks at the end of the previous subsection also apply here. Eq.~(\ref{Jdiamond}) gives in fact the modular conjugation of any region whose null projections coincide with those of the causal diamond. These regions have the same causal complement as the causal diamond, so Haag duality is satisfied for them also.

\section{Modular conjugation for multicomponent regions}\label{section6}

In this section we compute the modular conjugation for a generic open spacetime region ${\mathcal U}$. Unlike in the previous section, here it is crucial to restrict to the massless fermion in $1+1$, because it is only in this model that the relevant modular flow is known. The null projections of ${\mathcal U}$ will be collections of intervals,
\begin{equation}
    \pi^\pm({\mathcal U})=\bigcup_{i=1}^{n_\pm}\,(a_{i}^\pm,b_{i}^\pm).
\end{equation}
In what follows, for notational simplicity, we drop all subscripts and superscripts $\pm$; all the equations below hold for both chiralities. The modular flow is given by \cite{Erdmenger:2020nop}
\begin{equation}\label{modflowgen}
    \Delta^{is}\psi(x)\Delta^{-is} = 2\sinh(\pi s) \sum_{i=1}^{n} \frac{1}{\omega'(x_i(s))} \frac{1}{x - x_i(s)} \psi(x_i(s))\equiv\alpha_s(\psi(x)),
\end{equation}
where
\begin{equation}\label{omega}
    \omega(x) = \log\left(-\prod_{i=1}^{n}\frac{x-a_i}{x-b_i}\right)\qquad \omega(x_i(s)) = \omega(x) - 2\pi s.
\end{equation}
Note that $x$ lies in one of the intervals, because it is a null projection of a point in ${\mathcal U}$; this implies that $\omega(x)$ is real. Note also that, within each interval, $\omega$ is a monotonic function which goes from $-\infty$ at the left end of the interval to $+\infty$ at the right end. In consequence, the equation $\omega(y)=\omega(x)-2\pi s$ has exactly one solution $y$ in each interval; $x_i(s)$ denotes the solution that lies in the $i$-th interval. Unlike the modular flows we encountered in the previous section, the above modular flow is non-local: the modular evolution of a field operator localized in one of the components of ${\mathcal U}$ is a linear combination of field operators localized in each of the components.

\begin{figure}[ht]
    \centering
    \captionsetup{justification=centering}
    \includegraphics[width=0.5\textwidth]{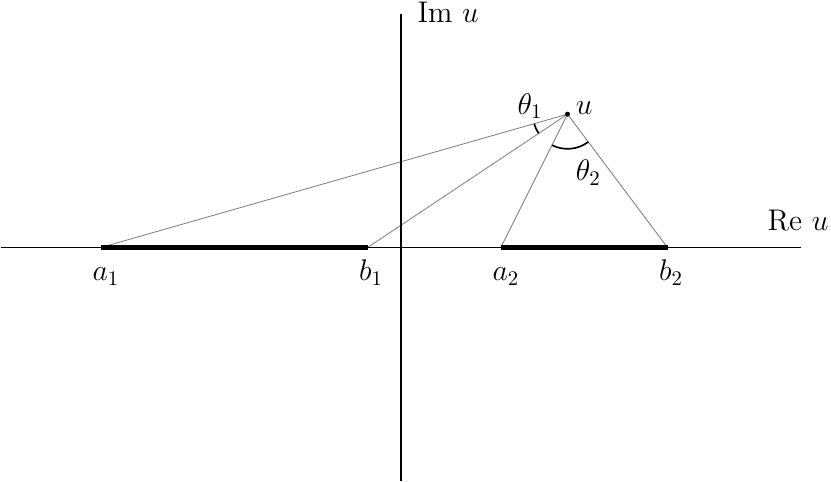}
    \caption{For $u$ in the upper half-plane, ${\text{arg}}(u-b_i)-{\text{arg}}(u-a_i)=\theta_i$, where $\theta_i$ is the angle shown in the figure. Clearly, $\sum_{i}\theta_i\in(0,\pi)$ and therefore $\sum_{i}[{\text{arg}}(u-a_i)-{\text{arg}}(u-b_i)]\in(-\pi,0)$.}
    \label{fig:imomega}
\end{figure}

\begin{figure}
    \centering
    \captionsetup{justification=centering}
    \includegraphics[scale=0.6]{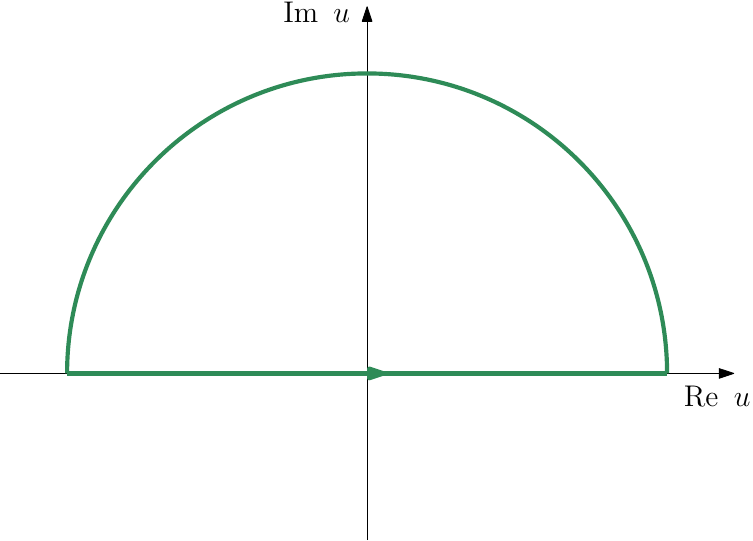}
    \caption{Integration contour for Eq.~(\ref{phix}).}
    \label{fig:contour}
\end{figure}

The action of the operator (\ref{modflowgen}) on the vacuum can be written in terms of a contour integral. Let us think of $\omega$ as a function on the complex plane, which has a cut on the complement of $\pi({\mathcal U})$ in the real line because the argument of the logarithm is negative there. For $u\in{\mathbb C}$, the imaginary part of $\omega(u)$ is
\begin{equation}
    \im \omega(u)=\sum_{i=1}^n\left[{\text{arg}}(u-a_i)-{\text{arg}}(u-b_i)\right]+(2k+1)\pi,
\end{equation}
where $k$ is an integer to be chosen so that $\im\omega(u)\in(-\pi,\pi]$. In figure \ref{fig:imomega} we show by a simple geometrical argument that, for $u$ in the upper half-plane, the above sum of $n$ terms lies in $(-\pi,0)$ and hence $\im\omega(u)\in(0,\pi)$. A similar argument shows that $\im\omega(u)\in(-\pi,0)$ for $u$ in the lower half-plane. Now, consider the vector-valued function
\begin{equation}\label{phix}
    |\phi_x(z)\rangle=-\frac{1}{2\pi i}\ointctrclockwise du\,\frac{\sinh\left[\frac{\omega(x)-\omega(u)}{2}\right]}{\sinh\left[\frac{\omega(x)-\omega(u)}{2}-\pi z\right]}\frac{1}{x-u}\psi(u)|\Omega\rangle
\end{equation}
for $\im z\in(-1/2,0)$, where the contour of integration is depicted in figure \ref{fig:contour}. Note that, despite the cuts of $\omega$, the ratio of hyperbolic sines above is meromorphic in $u$ (i.e., analytic except for poles). Recalling that $\psi(u)|\Omega\rangle$ is analytic in the upper half-plane and continuous on its boundary, it follows that the integral can be computed by residues. Since $\im\omega(u)\in(-\pi,\pi]$, the hyperbolic sine in the denominator only vanishes where its argument vanishes, so the only poles of the integrand are the $n$ solutions $u=x_i(z)$ of the equation $\omega(u)=\omega(x)-2\pi z$. By the discussion above, given the range of values of $z$, these all lie in the upper half-plane, i.e., within the contour. Applying the residue formula one obtains
\begin{equation}\label{phixmod}
    |\phi_x(z)\rangle=2\sinh(\pi z) \sum_{i=1}^{n} \frac{1}{\omega'(x_i(z))} \frac{1}{x - x_i(z)} \psi(x_i(z))|\Omega\rangle.
\end{equation}
Note from (\ref{phix}) that $|\phi_x(z)\rangle$ is an analytic (i.e., differentiable) function of $z$ everywhere in its domain, the strip $\im z\in(-1/2,0)$. Eq.~(\ref{phix}) would not make sense for $z$ on the boundary of that strip because in that case the contour would hit a pole, but still one can extend $|\phi_x\rangle$ to the boundary of the strip by continuity. Doing that, it follows from the above equation that, for $s\in{\mathbb R}$,
\begin{equation}
    |\phi_x(s)\rangle=\alpha_s(\psi(x))|\Omega\rangle.
\end{equation}
This gives the desired expression of the operator (\ref{modflowgen}) in terms of a contour integral. We will denote as $|\varphi_x(z)\rangle$ the function defined in the same way as $|\phi_x(z)\rangle$ but with $\psi$ replaced by $\psi^\dagger$; note that it has analogous properties, and in particular satisfies $|\varphi_x(s)\rangle=\alpha_s(\psi^\dagger(x))|\Omega\rangle$.

\begin{figure}[t]
    \centering
    \captionsetup{justification=centering}
    \includegraphics[scale=0.6]{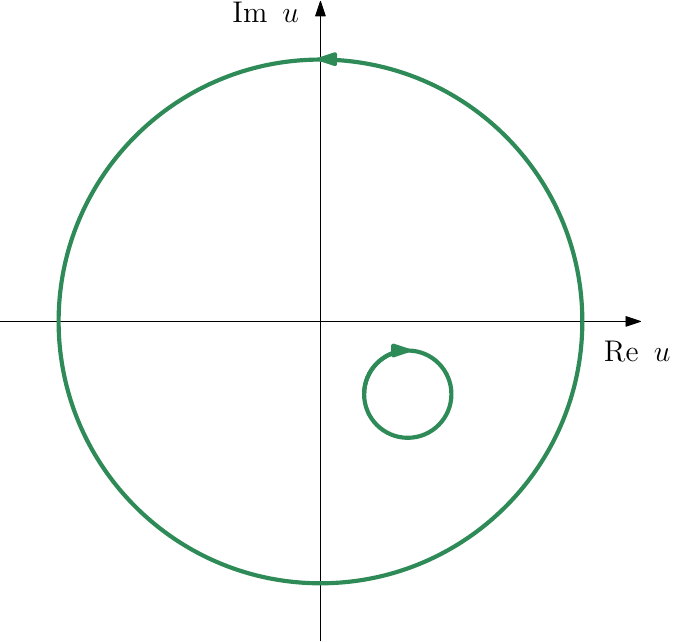}
    \caption{Integration contour for Eq.~(\ref{Gintegral}). The inner loop encircles the point $u=x-i\epsilon$.}
    \label{fig:contour2}
\end{figure}

Eq.~(\ref{modflowgen}) was derived in \cite{Erdmenger:2020nop} by a method based on the reduced density matrix, which strictly speaking is not well-defined in QFT. This is not really a problem, because one can make sense of the method by discretizing the theory and then sending the lattice spacing to zero at the end of the calculation. Still, it is desirable to verify the result by checking that $\alpha$ satisfies the modular condition. To do this, we compute for $\im z\in(-1/2,0)$
\begin{alignat}{2}\label{Gintegral}
    G(x,y;z)&\equiv\langle\Omega|\psi(x)|\varphi_y(z)\rangle=\langle\Omega|\psi^\dagger(x)|\phi_y(z)\rangle\nonumber\\
    &=-\left(\frac{1}{2\pi i}\right)^2\ointctrclockwise du\,\frac{\sinh\left[\frac{\omega(y)-\omega(u)}{2}\right]}{\sinh\left[\frac{\omega(y)-\omega(u)}{2}-\pi z\right]}\frac{1}{y-u}\,\frac{1}{x-u-i\epsilon}.
\end{alignat}
Note that the integrand above is analytic outside the contour of figure \ref{fig:contour} except for a pole at $u=x-i\epsilon$. Therefore, the contour can be deformed to that of figure \ref{fig:contour2}. The integral along the outer loop vanishes, because the integrand decays quickly at infinity, so one is left only with the contribution from the inner loop, which is determined by the pole at $u=x-i\epsilon$. The result is
\begin{equation}\label{G}
    G(x,y;z)=\frac{1}{2\pi i}\,\frac{\sinh\left[\frac{\omega(x)-\omega(y)}{2}\right]}{\sinh\left[\frac{\omega(x)-\omega(y)}{2}+\pi z\right]}\,\frac{1}{x-y}.
\end{equation}
Note that this function can be analytically continued to the strip $\im z\in(-1,0)$ and, after smearing, extends continuously to the boundary of that strip. By construction, it is clear that
\begin{equation}
    G(x,y;s-i\epsilon)=\langle\Omega|\psi(x)\alpha_s(\psi^\dagger(y))|\Omega\rangle=\langle\Omega|\psi^\dagger(x)\alpha_s(\psi(y))|\Omega\rangle.
\end{equation}
Moreover, it follows from (\ref{G}) that $G(x,y;s-i+i\epsilon)=G(y,x;-s-i\epsilon)$, so the modular condition is indeed satisfied, as we wanted to show.

\begin{figure}[ht]
    \centering
    \captionsetup{justification=centering}
    \includegraphics[width=0.4\textwidth]{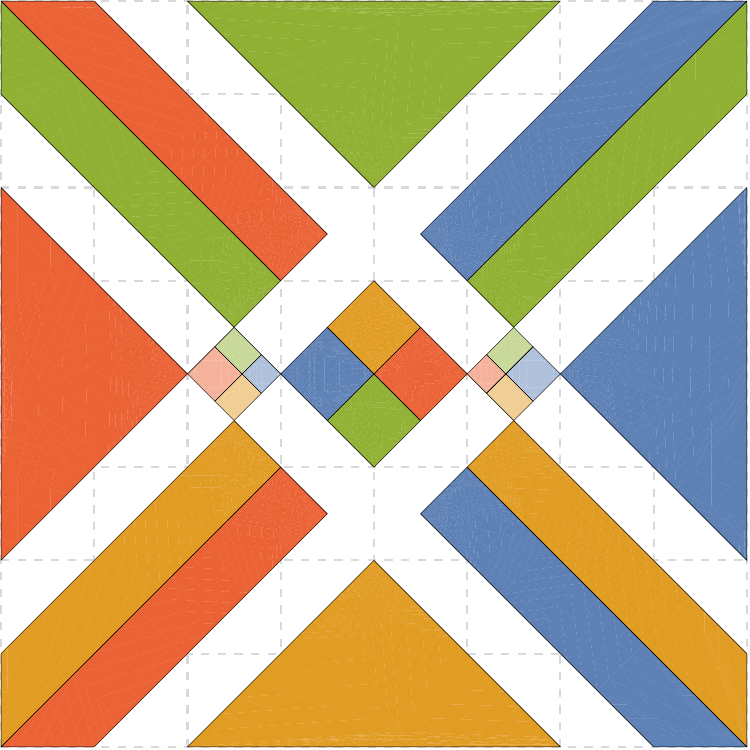}
    \caption{If ${\mathcal U}$ consists of two spacelike separated and symmetrically arranged causal diamonds, regions of ${\mathcal U}$ (lighter colored) are mapped to regions of the corresponding darker color under the maps $x^\pm\mapsto\bar x_{1,2}^\pm$.}
    \label{Fig:J2interval}
\end{figure}

Let us now compute the modular conjugation. We proceed analogously to the case of the Rindler wedge, section \ref{section4}, by analytic continuation from the modular flow. From (\ref{modflowgen}) we have
\begin{equation}
     \Delta^{is}\psi^\dagger(x)|\Omega\rangle=2\sinh(\pi s) \sum_{i=1}^{n} \frac{1}{\omega'(x_i(s))} \frac{1}{x - x_i(s)} \psi^\dagger(x_i(s))|\Omega\rangle.
\end{equation}
As we know, the left-hand side is analytic in $s$ on the interior of the strip $\im s\in[-1/2,0]$ and continuous on its boundary. And, by the discussion around (\ref{phixmod}), the same is true for the right-hand side. Note that the terms of the sum on the right-hand side are not separately analytic, because $x_i(s)$, which is a root of a polynomial of degree $n$, generically has branch cuts (think for example of the case $n=2$, where the formula for $x_i(s)$ involves a square root). But the sum of all terms is analytic, because it can be rewritten in the explicitly analytic form (\ref{phix}), (\ref{phixmod}) (with $\psi$ replaced by $\psi^\dagger$). Therefore, the above equation holds everywhere in the strip $\im s\in[-1/2,0]$, and in particular for $s=-i/2$,
\begin{equation}
    \Delta^{1/2}\psi^\dagger(x)|\Omega\rangle=-2i\sum_{i=1}^n\frac{1}{\omega'(\bar x_i)}\frac{1}{x-\bar x_i}\psi^\dagger(\bar x_i)|\Omega\rangle,
\end{equation}
where 
\begin{equation}
    \omega(\bar x_i)=\omega(x)+i\pi.
\end{equation}
Note that $\bar x_i$ lies in the complement of $\pi({\mathcal U})$ in the real line, which means that $Z\psi^\dagger(\bar x_i)Z^\dagger\in{\mathcal A}_{\mathcal U}'$. With this in mind, and following the same steps as in the case of the Rindler wedge, Eqs.~(\ref{delta12rind})-(\ref{Jpsi}), we obtain
\begin{equation}\label{eljota}
    J\psi(x)J=2\sum_{i=1}^n\frac{1}{\omega'(\bar x_i)}\frac{1}{x-\bar x_i}Z\psi^\dagger(\bar x_i)Z^\dagger.
\end{equation}
This is the main result of the paper. The maps $x^\pm\mapsto\bar x_{i}^\pm$ are represented in figure \ref{Fig:J2interval} in the case of a simple choice of the region ${\mathcal U}$. 
Note from this equation (and also from the figure) that ${\mathcal A}_{\mathcal U}'=J{\mathcal A}_{\mathcal U}J\subseteq{Z\mathcal A}_{\mathcal U^*}Z^\dagger$, where, again, ${\mathcal U}^*$ denotes the null complement of ${\mathcal U}$, i.e., the the largest open set of points which cannot be connected with ${\mathcal U}$ by a null line. Since we already know that the opposite inclusion holds, we conclude that
\begin{equation}\label{commutant}
    {\mathcal A}_{\mathcal U}'=Z{\mathcal A}_{\mathcal U^*}Z^\dagger.
\end{equation}
Thus, whether Haag duality holds or not for a region ${\mathcal U}$ depends entirely on whether the null projections of ${\mathcal U}'$ coincide with those of ${\mathcal U}^*$ (which form the complement of the projections of ${\mathcal U}$). We will discuss this in more detail below. Another feature to highlight about Eq.~(\ref{eljota}) is that, unlike the previous cases discussed, the map induced by $J$ on operators is non-local. This arises, of course, due to the non-locality of the modular flow. As a simple check of (\ref{eljota}), setting $n=1$ and $b=-a=R$ one recovers exactly the result for the causal diamond, Eq.~(\ref{Jdiamond}), as it should be. We have checked that the result for the causal diamond is also recovered in more complicated ways, for example starting with two intervals and making them approach one another or sending one to infinity.

\subsection{Haag duality}

As is clear from (\ref{commutant}), in the model we are considering Haag duality is satisfied for a region ${\mathcal U}$ if and only if ${\mathcal A}_{{\mathcal U}'}={\mathcal A}_{{\mathcal U}^*}$, which is equivalent to saying that $\pi^\pm({\mathcal U}')=\pi^\pm({\mathcal U}^*)$ or, in other words, that the null projections of ${\mathcal U}'$ form the complement of those of ${\mathcal U}$ in the null axes. Let us see some examples of regions for which Haag duality holds, and others for which it fails.

\begin{figure}[ht]
    \centering
    \captionsetup{justification=centering}
    \includegraphics[scale=0.6]{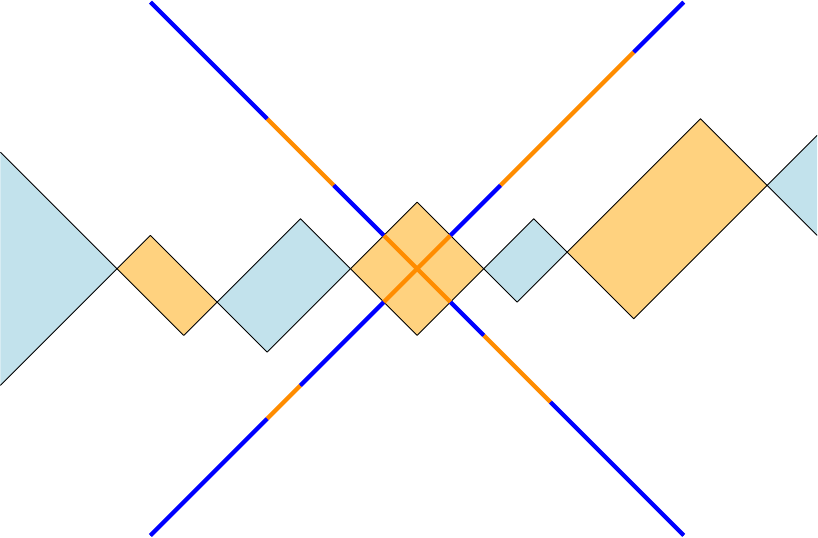}
    \caption{A generic example of a causally complete region (orange) and its causal complement (blue). The null projections of the region and its causal complement are the collections of segments of the corresponding color. These fill the null axes, which means that Haag duality holds for this region.}
    \label{fig:causallycomplete}
\end{figure}

A region ${\mathcal U}$ is said to be {\emph{causally complete}} if it satisfies ${\mathcal U}''={\mathcal U}$; for ${\mathcal U}$ generic, ${\mathcal U}''$ is called the {\emph{causal completion}} of ${\mathcal U}$. Note that ${\mathcal U}''$ always contains ${\mathcal U}$ and has the same causal complement. Hence we have 
\begin{equation}
    Z{\mathcal A}_{{\mathcal U}'}Z^\dagger\subseteq {\mathcal A}_{{\mathcal U}''}'\subseteq{\mathcal A}_{\mathcal U}',
\end{equation}
so, if Haag duality holds for ${\mathcal U}$, then it also holds for its causal completion. Causal completions are always causally complete, so, for this reason, Haag duality is more likely to hold for causally complete regions. In figure \ref{fig:causallycomplete} we show a generic example of a causally complete region. Its null projections and those of its causal complement fill the null axes, so Haag duality is satisfied for this region. We expect this to remain true for arbitrary causally complete regions.

\begin{figure}[ht]
    \centering
    \captionsetup{justification=centering}
    \includegraphics[scale=0.6]{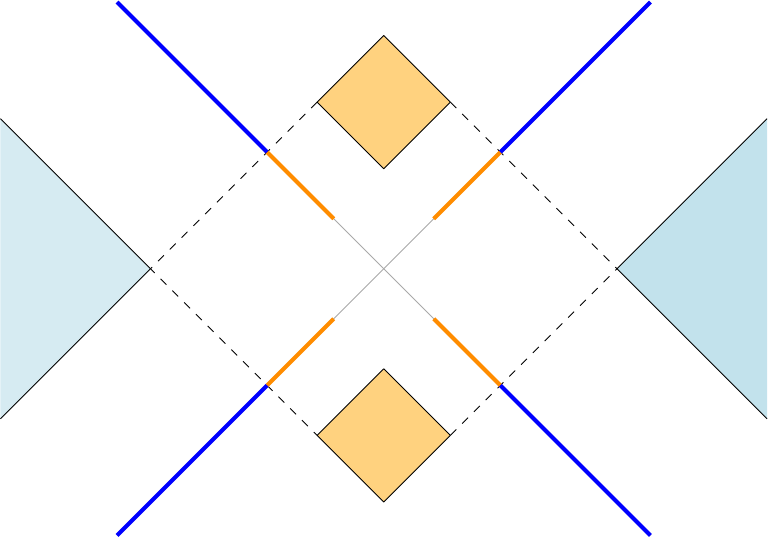}
    \caption{A non-causally complete region (orange) and its causal complement (blue). The null projections of the region and its causal complement are the collections of segments of the corresponding color. The central portion of the null axes is not painted, which means that Haag duality fails for this region.}
    \label{fig:noncc}
\end{figure}

In figure \ref{fig:noncc} we show an example of a non-causally complete region. Its null projections and those of the causal complement do not fill the null axes, so Haag duality does not hold for this region. This is not generic of non-causally complete regions: in section \ref{s4} we have already seen examples of non-causally complete regions for which Haag duality holds: these are the blue regions of figures \ref{figprojections} and \ref{figrindler}.

\section{Final comments}\label{section7}

In this paper we computed a new modular conjugation, namely that of a $1+1$ dimensional free massless Dirac field in the vacuum state for multicomponent regions. We also revisited some previously known results: modular conjugation for the vacuum of any QFT in the Rindler wedge (Bisognano-Wichmann) and for the vacuum of any CFT in the ball (Hislop-Longo), with emphasis on completing the details that arise when one considers fermions instead of bosons.

To compute the modular conjugation we first studied the action of the modular flow on the fields applied to the cyclic and separating vacuum state. Then, we extended this to modular parameter $s=-i/2$ in order to relate it with the action of $J$ on the fields when applied to the vacuum. Finally, using the separating property of the vacuum state we were able to get rid of the vacuum state and obtained an operator equation relating the action of $J$ with the modular flow.

For the simple cases of Bisognano-Wichmann \cite{Bisognano:1976za} and Hislop-Longo \cite{Hislop:1981uh}, the modular flow is local and $J$ acts geometrically. In the novel case of multicomponent regions the modular flow is known to be non-local and exhibits a ``mixing'' between components \cite{Casini_2009}.
This translates into non-localities in the modular conjugation as well.

For the massless free Dirac field in $1+1$ dimensions the result for $J$ for arbitrary regions gives us a large playground to test the validity of Haag duality. We observed that Haag duality holds in this model for the local algebras associated to generic causally complete regions. When the regions are not causally complete we found situations in which duality holds but others in which it does not. We hope that the results found here in this regard might be helpful in the task of elucidating under which conditions one can expect this property to hold in a general QFT.

Notice that in order to obtain the modular conjugation we heavily relied on the knowledge of the modular flow. Therefore, one could in principle obtain $J$ for other models in which the modular flow has been studied. This is for example the case of the free massless Dirac field but on the circle at non-zero temperature \cite{Blanco:2019xwi,Blanco:2019cet,Fries:2019ozf}.

The modular conjugation for a two-component region for the massless fermion in $1+1$ dimensions was recently considered in \cite{Mintchev:2022fcp}.
In that work, a connection between the action of some modular conjugations and the geodesic bit threads of the corresponding dual gravitational backgrounds was also made. The result for the new multicomponent modular conjugation we provided in this manuscript might perhaps be useful to explore that interesting connection further.

\section*{Acknowledgements}

We thank Horacio Casini and Alan Garbarz for useful discussions. This work has been partially supported by CONICET, UBA and through the grant UBACyT 20020190200162BA. DB acknowledges support from the ANPCyT through the grant PICT-2018-03593.

\newpage

\printbibliography

\end{document}